\documentclass[11pt]{article}
\usepackage{amsmath,amssymb}
\usepackage{graphicx}
\usepackage[table]{xcolor}           

\usepackage{makecell}

\usepackage{array}
\usepackage{booktabs}
\usepackage{siunitx}
\usepackage[utf8]{inputenc}
\usepackage{multirow}
\usepackage{fmtcount}
\usepackage{caption}
\usepackage[colorlinks=true, 
            linkcolor=blue,      
            citecolor=blue,      
            urlcolor=blue]{hyperref}   
\usepackage{fancyhdr}
\usepackage{cleveref}
\usepackage{slashed}
\usepackage{dcolumn}
\usepackage{bm}
\usepackage{rotating}
\usepackage{lscape}
\usepackage{cite}
\usepackage{float}
\usepackage{url}
\usepackage{mathtools}
\usepackage{physics}
\usepackage{mathrsfs}
\usepackage{simplewick}
\usepackage{tikz}
\usepackage{subcaption}

\renewcommand{\arraystretch}{1.8}

\usetikzlibrary{arrows,shapes,trees,matrix,positioning,calc,through,
                decorations.pathreplacing,decorations.pathmorphing,
                decorations.markings}

\newcolumntype{C}[1]{>{\centering\arraybackslash}m{#1}}

\newcommand{\subtitle}[1]{%
    \posttitle{%
        \par\end{center}
        \begin{center}\large#1\end{center}
        \vskip0.5em}%
}

\def\Re{{\cal R \mskip-4mu \lower.1ex \hbox{\it e}\,}}
\def\Im{{\cal I \mskip-5mu \lower.1ex \hbox{\it m}\,}}
\def\to{\rightarrow}
\def\beq{\begin{equation}}
\def\eeq{\end{equation}}
\def\bea{\begin{eqnarray}}
\def\eea{\end{eqnarray}}

\def\sla#1{\raise.15ex\hbox{$/$}\kern-.57em #1}

\begin{document}

\begin{center}
\vspace*{15mm}
\vspace{1cm}
{\Large \bf W-boson helicity fractions in top decay as probes of dimension-6 and dimension-8 SMEFT operators}

\vspace{1cm}

{\bf Afsaneh Kianfar, Gholamhossein Haghighat and Mojtaba Mohammadi Najafabadi}\\
{\small\sl 
School of Particles and Accelerators, Institute for Research in Fundamental Sciences (IPM) P.O. Box 19395-5531, Tehran, Iran\\ }
\vspace*{.2cm}
\end{center}

\vspace*{10mm}

\begin{abstract}
Precision measurements of top-quark decays provide powerful probes of physics beyond the Standard Model (SM). 
While the impact of dimension-6 operators in the SM Effective Field Theory (SMEFT) 
has been extensively studied, the role of dimension-8 contributions remains largely unexplored, despite their 
potential importance as experimental precision improves. In this work, we present a combined 
analysis of dimension-6 and a representative subset of dimension-8 SMEFT effects using the W-boson helicity fractions in top-quark decays. 
We compute the leading-order contributions of these operators to the top-quark decay width and helicity fractions, 
and perform one-parameter and selected two-parameter $\chi^{2}$ fits to the combined ATLAS and CMS measurements 
at a reference scale $\Lambda=1$~TeV. 
From the fit results, we find that the inclusion of dimension-8 contributions 
affects the allowed parameter space of several dimension-6 coefficients through non-trivial correlations and degeneracies. 
Since the leading dimension-8 contributions enter at the same order $\mathcal{O}(\Lambda^{-4})$ as the squared dimension-6 terms retained 
in our analysis, this highlights the importance of a consistent treatment of the EFT expansion when interpreting SMEFT constraints.
\end{abstract}

\newpage
\section{Introduction}

Despite many successes of the SM, it  leaves several observational and theoretical puzzles unresolved. 
These include the nature of dark matter, the origin of neutrino masses, 
the baryon asymmetry of the universe, and persistent theoretical challenges such as the hierarchy problem. 
A wide range of theories beyond the SM (BSM) have been proposed to address these shortcomings. 
However, extensive searches at the Large Hadron Collider (LHC) have yielded no direct evidence for new particles at the TeV scale. 
This absence of direct discoveries has intensified the focus on indirect probes of new physics, where precision measurements of SM 
processes are used to detect the subtle effects of heavy or weakly-coupled new states.\\
The SMEFT provides a powerful and model independent framework for such precision 
studies~\cite{Buchmuller:1985jz, Grzadkowski:2010es}. SMEFT extends the SM Lagrangian by adding a series 
of higher-dimensional operators, constructed from SM fields and respecting its gauge symmetries, 
suppressed by powers of the new physics scale $\Lambda$. The leading deviations from the SM 
are parameterized by dimension-6 operators, while dimension-8 operators encode the next-to-leading corrections. 
Although formally more suppressed, dimension-8 effects can become
phenomenologically relevant when SMEFT analyses are interpreted
beyond the leading $\mathcal{O}(\Lambda^{-2})$ level. In particular,
interference between the SM and dimension-8 operators contributes
at $\mathcal{O}(\Lambda^{-4})$, the same order as squared
dimension-6 terms, and can therefore affect the interpretation of
dimension-6 fits~\cite{Brivio:2017vri}.
A consistent connection between low-energy observables and underlying new physics effects is established through matching between 
effective field theories, such as SMEFT above the electroweak scale and low-energy effective field theory (LEFT) below it. 
In this context, higher-dimensional contributions beyond dimension six, in particular dimension-eight operators, 
become relevant for a consistent and precise interpretation of experimental data \cite{Hamoudou:2022dwy}.\\
Top-quark physics offers a uniquely sensitive laboratory for testing the SMEFT framework. As the heaviest known elementary particle, 
the top quark has the largest coupling to the Higgs sector and decays almost exclusively via $t \to W b$ \cite{beneke, Bernreuther:2008ju}. 
This makes precision studies of this decay channel a direct probe of effective operators that modify the $Wtb$ vertex. 
Among the most sensitive observables are the W-boson helicity fractions; $F_0$ (longitudinal), $F_L$ (left-handed), 
and $F_R$ (right-handed), which are extracted from the angular distributions of decay leptons in semileptonic top decays. 
Recent LHC measurements and combinations of W-boson helicity fractions in top-quark decays have reached percent-level precision~\cite{CMS:2023wjg,Aad:2020px,ATLAS:2022vfh},
establishing them as powerful constraints for indirect BSM searches.
The sensitivity of the $W$-boson helicity fractions in top-quark decay have been investigated in several beyond the SM scenarios, 
such as the two-Higgs-doublet model, top-color assisted technicolor (TC2), and the noncommutative extension of the SM~\cite{mmn1, mmn2}.\\
The impact of dimension-6 SMEFT operators on the decay $t \to Wb$ and $Wtb$ vertex have been investigated in considerable detail, 
with next-to-leading-order QCD calculations now available, including the effects of four-fermion interactions and dipole 
operators~\cite{Zhang:2010dr, Boughezal:2019kfp, Zhang:2012cd, Buckley:2015lku, Hartland:2019bjb, Gao:2014rja, r1, r2, r3, r4, r5, r6, r7,r8,r9,r10}. 
These studies provide a mature understanding of the leading deviations from the SM in top-quark decay.
In contrast, the phenomenological impact of dimension-8 operators
in top-quark decay remains comparatively unexplored. 
Global SMEFT analyses that truncate the expansion at dimension-6 may lead to incomplete or biased interpretations of data, 
because effects from dimension-8 operators can imitate, distort, or partly screen the contributions attributed to 
dimension-6 interactions~\cite{Banerjee:2019twi, Alioli:2022oxm}. As the precision of LHC Run~3 measurements 
improves and the High-Luminosity LHC enters operation, a consistent treatment of the EFT expansion demands that such subleading terms be incorporated.\\
While several works have examined aspects of the dimension-8 SMEFT framework \cite{Gillies:2024xxx, Flores-Hernandez:2026sac, d80,d81,d82,d83,d84,d85,d86}, they do not provide a decay-level 
phenomenological analysis of $t \to Wb$. The structure and classification of dimension-8 operators have been studied 
in detail in the literature. Broader investigations have emphasized the formulation of operator bases, the consistency of 
the effective field theory framework, and multi-process considerations~\cite{Murphy:2020rsh}. The specific dynamics of top-quark 
decay are therefore largely absent from the existing dimension-8 literature. 
The focus of this work is therefore the decay $t \to Wb$. We consider
dimension-6 SMEFT operators together with a representative subset of
dimension-8 operators that contribute to the on-shell $Wtb$ vertex at
tree level. 
Within this setup we retain the interference between the
SM and dimension-6 operators, the squared
dimension-6 terms, and the interference between the
SM and the dimension-8 operators,
consistently keeping all contributions up to
$\mathcal{O}(\Lambda^{-4})$.
We compute the leading-order contributions to the decay width and helicity fractions
within the assumptions of an on-shell $t\to Wb$ decay, and we perform one- and two-parameter
fits to extract constraints on the corresponding Wilson coefficients.
This setup allows us to quantify how dimension-8 effects can modify the interpretation of
dimension-6 fits in the top-quark sector.
\medskip
The remainder of this paper is organized as follows. 
In Sec.~\ref{sec:framework}, we introduce the SMEFT framework and the set of dimension-6 and representative dimension-8 operators relevant for $t \to W b$ decay. 
In Sec.~\ref{sec:calculation}, we present the calculation of the decay width and the $W$-boson helicity fractions, including all contributions up to $\mathcal{O}(\Lambda^{-4})$. 
In Sec.~\ref{secres}, we describe the statistical analysis and present the results of the one- and two-parameter fits. 
The physical implications of these results are discussed in Sec.~\ref{secdis}. 
Finally, we summarize our findings and conclude in Sec.~\ref{secsummary}.


\section{Theoretical framework}
\label{sec:framework}

In this section, we introduce the SMEFT framework adopted in our analysis and specify the subset of dimension-6 and dimension-8 
operators relevant for the decay $t \to W b$. We then relate these operators to the effective $Wtb$ vertex form factors that enter the calculation of the decay width and helicity fractions.

\subsection{SMEFT Framework and Operator Basis}

The SMEFT provides a systematic framework to parameterize potential deviations from the SM arising from new physics at a high scale $\Lambda$. 
The SMEFT Lagrangian extends the SM Lagrangian with higher-dimensional operators constructed from SM fields and 
respecting the SM gauge symmetries $SU(3)_C \times SU(2)_L \times U(1)_Y$.
The complete Lagrangian is organized as an expansion in inverse powers of the new physics scale \cite{Buchmuller:1985jz}:
\begin{equation}
\mathcal{L}_{\text{SMEFT}} = \mathcal{L}_{\text{SM}} + \sum_{n=6}^{\infty} \frac{1}{\Lambda^{n-4}} \sum_k C_k^{(n)} Q_k^{(n)}.
\label{eq:smeft_lagrangian}
\end{equation}
Here $Q_k^{(n)}$ are dimension-$n$ operators and $C_k^{(n)}$ the corresponding Wilson coefficients. The coefficients are taken to be dimensionless, 
with the scale suppression made explicit by $1/\Lambda^{n-4}$.
The unique gauge-invariant dimension-5 operator built from SM fields is the Weinberg operator $(LH)(LH)$, which violates lepton number. ‌‌
Because we restrict to baryon- and lepton-number conserving interactions, our expansion begins at dimension~6:
\begin{equation}
\mathcal{L}_{\text{SMEFT}} = \mathcal{L}_{\text{SM}} + \frac{1}{\Lambda^2} \sum_i C_i^{(6)} Q_i^{(6)} + \frac{1}{\Lambda^4} \sum_j C_j^{(8)} Q_j^{(8)} + \mathcal{O}\left(\frac{1}{\Lambda^6}\right).
\end{equation}
In this work, we consistently include all contributions to the observables up to $\mathcal{O}(1/\Lambda^4)$ in the EFT expansion. 
This comprises three classes of terms: 
(i) the interference between the SM and dimension-6 operators $\big(\mathcal{O}(1/\Lambda^2)\big)$, 
(ii) the square of dimension-6 amplitudes $\big(\mathcal{O}(1/\Lambda^4)\big)$, and 
(iii) the interference between the SM and dimension-8 operators $\big(\mathcal{O}(1/\Lambda^4)\big)$. 
Contributions from dimension-6--dimension-8 interference $\big(\mathcal{O}(1/\Lambda^6)\big)$ 
and pure dimension-8 squares $\big(\mathcal{O}(1/\Lambda^8)\big)$ are formally of higher order 
and are neglected in our analysis, consistent with a truncation at next-to-leading order in the 
EFT expansion. This truncation scheme is applied uniformly to all observables considered.

\subsection{Relevant Operators for $t \to W b$ Decay }

After electroweak symmetry breaking (EWSB), the Higgs doublet acquires a vacuum expectation value
\begin{equation}
\langle H \rangle = \frac{1}{\sqrt{2}} \begin{pmatrix} 0 \\ v \end{pmatrix}, \qquad v \approx 246~\text{GeV},
\end{equation}
which generates masses for the $W$, $Z$, and fermions. The physical gauge bosons are related to gauge eigenstates by
\begin{equation}
W^\pm_\mu = \frac{1}{\sqrt{2}}(W^1_\mu \mp i W^2_\mu), \quad 
\begin{pmatrix} Z_\mu \\ A_\mu \end{pmatrix} = 
\begin{pmatrix} c_W & -s_W \\ s_W & c_W \end{pmatrix}
\begin{pmatrix} W^3_\mu \\ B_\mu \end{pmatrix},
\end{equation}
where $c_W=\cos\theta_W$, $s_W=\sin\theta_W$, and $\theta_W$ is the weak mixing angle.
Quark mass eigenstates are obtained via unitary rotations of the weak eigenstates. The misalignment of the up- and down-type rotations gives 
the Cabibbo–Kobayashi–Maskawa (CKM) matrix, $V_{\text{CKM}} = U_u^\dagger U_d$, which appears in the charged-current interactions.
When considering higher-dimensional operators, redundancies can be removed using equations of motion and field redefinitions. 

The decay $t \to W b$ proceeds through charged-current interactions that are modified by SMEFT operators. 
At tree level, the relevant operators are those that directly affect the $Wtb$ vertex or induce new Lorentz structures. We classify them by mass dimension and their transformation properties under CP.
We work in the Warsaw basis for dimension-6 operators \cite{Grzadkowski:2010es, AguilarSaavedra:2006fy} and adopt a consistent extension for the subset of dimension-8 operators relevant to $t\to Wb$ \cite{Murphy:2020rsh,Li:2020gnx}.
The most general on-shell $Wtb$ interaction can be parameterized as
\begin{equation}
\Gamma_{tWb}^\mu = -\frac{g}{\sqrt{2}}\left[\gamma^\mu\big(f_V^L P_L + f_V^R P_R\big) + \frac{i\sigma^{\mu\nu}}{m_W}(p_t - p_b)_\nu \big(f_T^L P_L + f_T^R P_R\big)\right],
\label{eq:Gamma}
\end{equation}
with $P_{L,R} = (1 \mp \gamma_5)/2$. In the SM at tree level: $f_V^L = V_{tb},  f_V^R = f_T^L = f_T^R = 0.$
The form factors $f_V^{L,R}$ and $f_T^{L,R}$ are defined to be dimensionless quantities by construction. Therefore, appropriate normalization factors 
have been explicitly included in the matching relations to ensure dimensional consistency of the effective $Wtb$ vertex parametrization. The SMEFT operators modify these form factors as
\begin{align}
f_V^L &= V_{tb} + \frac{v^2}{\Lambda^2}\,\Delta f_V^{L,(6)} + \frac{v^4}{2\Lambda^4}\,\Delta f_V^{L,(8)} + \cdots, \\
f_V^R &= \frac{v^2}{2\Lambda^2}\,\Delta f_V^{R,(6)} + \frac{v^4}{2\Lambda^4}\,\Delta f_V^{R,(8)} + \cdots, \\
f_T^L &= \frac{v}{\Lambda^2}\,\Delta f_T^{L,(6)} + \frac{v^3}{2\Lambda^4}\,\Delta f_T^{L,(8)} + \cdots, \\
f_T^R &= \frac{v}{\Lambda^2}\,\Delta f_T^{R,(6)} + \frac{v^3}{2\Lambda^4}\,\Delta f_T^{R,(8)} + \cdots,
\end{align}
where the quantities $\Delta f_i^{(6)}$ and $\Delta f_i^{(8)}$ denote the shifts in the form factors induced by dimension-6 and dimension-8 operators, respectively. 
Their explicit expressions are linear combinations of the corresponding Wilson coefficients, weighted by electroweak parameters and CKM matrix elements. 
For clarity, we summarize the operator contributions to these shifts  in the following subsections.

\subsubsection{Dimension-6 Operators}

At tree level, four CP-even Warsaw-basis operators contribute directly to the $t \to Wb$ transition which are presented in Table \ref{tab:dim6_operators}. 
They modify either the left-handed charged current or induce right-handed vector and tensor structures.

\begin{table}[h!]
\centering
\renewcommand{\arraystretch}{1.3}
\begin{tabular}{@{}lll@{}}
\toprule
\textbf{Operator} & \textbf{Structure} & \textbf{CP} \\
\midrule
$Q_{Hq}^{(3)}$ & $(H^\dagger i\overleftrightarrow{D}{}_\mu^I H)(\bar{q}_p \tau^I \gamma^\mu q_r)$ & Even \\
$Q_{Hud}^{(3)}$      & $i(\tilde{H}^\dagger D_\mu H)(\bar{u}_p \gamma^\mu d_r)$ & Even \\ 
$Q_{uW}$       & $(\bar{q}_p \sigma^{\mu\nu} u_r)\,\tau^I \tilde{H}\, W_{\mu\nu}^I$ & Even \\
$Q_{dW}$       & $(\bar{q}_p \sigma^{\mu\nu} d_r)\,\tau^I H\, W_{\mu\nu}^I$ & Even \\
\bottomrule
\end{tabular}
\caption{Dimension-6 operators contributing to $t \to W b$ at tree level. 
The indices $p,r$ denote flavor generations.}
\label{tab:dim6_operators}
\end{table}

After electroweak symmetry breaking, these operators induce corrections to the $Wtb$ form factors $f_V^{L,R}$ and $f_T^{L,R}$, 
including CKM factors from quark mixing. These corrections have the following form:
\begin{align}
\Delta f_V^{L,(6)} &= V_{tb} \, C_{Hq}^{(3)}, \qquad \Delta f_V^{R,(6)} = C_{Hud}^{(3)}, \\
\Delta f_T^{L,(6)} &= \frac{2\sqrt{2}~m_W}{g}C_{dW}, \qquad \Delta f_T^{R,(6)} = \frac{2\sqrt{2}~m_W}{g} V_{tb} \, C_{uW}.
\end{align}

\subsubsection{Dimension-8 Operators}

At dimension~8, operators that map onto the vertex structures in Eq.~(\ref{eq:Gamma}) and reduce after EWSB
 to effective corrections contributing to $t \to Wb$. The representative subset considered in this work, 
 shown in Table~\ref{tab:dim8_operators}, includes both CP-even and CP-odd structures at the level of the formal operator basis.

\begin{table}[h!]
\centering
\renewcommand{\arraystretch}{1.3}
\begin{tabular}{@{}lll@{}}
\toprule
\textbf{Operator} & \textbf{Structure} & \textbf{CP} \\
\midrule
\multicolumn{3}{c}{\textbf{CP-even}} \\
\midrule
$Q_{q^2 H^4 D}^{(2)}$ & $ i(\overline q_p\gamma^\mu\tau^Iq_r)[(H^\dagger\overleftrightarrow D^I_\mu H)(H^\dagger H)+(H^\dagger\overleftrightarrow D_\mu H)(H^\dagger\tau^IH)]$ & Even \\
$Q_{ud H^4 D}$ & $i(\bar{u}\gamma^\mu d)\,(\tilde{H}^\dagger \overleftrightarrow{D}_\mu H)(H^\dagger H)$ & Even \\
$Q_{quWH^3}^{(1)}$    & $(\bar{q}\sigma^{\mu\nu} u)\,\tau^I \tilde{H}\,(H^\dagger H)\, W_{\mu\nu}^I$ & Even\\
$Q_{qdWH^3}^{(1)}$    & $(\bar{q}\sigma^{\mu\nu} d)\,\tau^I H\,(H^\dagger H)\, W_{\mu\nu}^I$ & Even \\ 
\midrule
\multicolumn{3}{c}{\textbf{CP-odd}} \\
\midrule
$Q_{q^2 H^4 D}^{(3)}$ & $i\epsilon^{IJK}(\bar{q}\gamma^\mu \tau^I q)(H^\dagger \overleftrightarrow{D}{}_\mu^J H)(H^\dagger \tau^K H)$ & Odd \\
\bottomrule
\end{tabular}
\caption{Representative CP-even and CP-odd dimension-8 operators relevant to $t \to W b$ decay.}
\label{tab:dim8_operators}
\end{table}

As discussed above, the inclusion of dimension-8 terms is essential for a consistent treatment of the 
EFT expansion when considering observables beyond leading order in $1/\Lambda$. 
In particular, the interference between the SM and dimension-8 operators 
contributes at $\mathcal{O}(\Lambda^{-4})$, the same order as squared dimension-6 terms, 
and can therefore modify correlations and degeneracy directions in the parameter space of Wilson coefficients.
The set of SMEFT operators retained in this analysis is defined by requiring non-vanishing tree-level contributions 
to the decay amplitude of $t \to W b$ through modifications of the on-shell $Wtb$ vertex. As shown in Table \ref{tab:dim6_operators}, at dimension~6, 
the operators $Q_{Hq}^{(3)}$, $Q_{Hud}^{(3)}$, $Q_{uW}$, and $Q_{dW}$ form the complete basis of tree-level contributions, 
generating all independent left-handed, right-handed, and dipole structures relevant for the decay rate and helicity fractions at $\mathcal{O}(\Lambda^{-2})$.
At dimension~8, we consider a representative subset of operators that contribute at tree level under the assumptions of a two-body decay and an on-shell $W$ boson. 
After EWSB, these operators induce corrections to the $Wtb$ vertex that scale as $\mathcal{O}(\Lambda^{-4})$. This selection is not exhaustive, but provides a 
consistent and phenomenologically relevant framework to study the impact of higher-dimensional effects on the extraction of dimension-6 coefficients. 
Complete classifications of dimension-8 SMEFT operators can be found in Refs.~\cite{Murphy:2020rsh,Li:2020gnx}.

CP-odd operators are included in the formal basis for completeness. However, the observables considered in this work, namely the $W$-boson helicity fractions, are CP-even quantities. 
As a result, the interference between the SM and CP-odd dimension-8 operators vanishes, and their leading contributions arise only at higher order. 
Since our analysis consistently retains terms up to $\mathcal{O}(\Lambda^{-4})$, these operators do not contribute to the observables considered here 
and therefore do not affect the numerical constraints.

Overall, the operator basis used in this work consists of nine SMEFT operators (four at dimension-6 and five at dimension-8) that can contribute to $t \to W b$ at tree level. 
However, not all coefficients are equally constrained by the available observables. In particular, the one-parameter scans for $C_{Hq}^{(3)}$ and $C_{q^2H^4D}^{(2)}$ 
exhibit approximate flat directions, preventing the extraction of finite confidence intervals from the present dataset. 
The operators $Q_{Hq}^{(3)}$ and $Q_{q^2H^4D}^{(2)}$ modify the same left-handed vector structure of the $Wtb$ vertex, differing only in their EFT order. 
Since they do not introduce new Lorentz structures, their effects on the $W$-boson helicity fractions largely cancel in the ratios, leading to these flat directions. 
Consequently, we do not report constraints on these operators in the present analysis.

After establishing the operator basis, we relate the corresponding Wilson coefficients to the effective $Wtb$ vertex form factors. 
The deviations induced by the dimension-8 operators can be expressed as
\begin{align}
\Delta f_V^{L,(8)} &= V_{tb}^{*} \left(C_{q^2 H^4 D}^{(2)}-\frac{i}{2} \, C_{q^2 H^4 D}^{(3)}\right), \quad \Delta f_V^{R,(8)} =  C_{udH^4D}, \\
\Delta f_T^{L,(8)} &=  \frac{2\sqrt{2}~m_W}{g} C_{qdWH^3}^{(1)}, \quad \Delta f_T^{R,(8)} = \frac{2\sqrt{2}~m_W}{g} V_{tb} \, C_{quWH^3}^{(1)}.
\end{align}
These expressions explicitly show how higher-dimensional operator structures modify the strength, chiral properties, and Lorentz structure of the 
charged-current interaction beyond the dimension-6 level.

\section{W-boson helicity observables}
\label{sec:calculation}

In the decay $t \to W^{+} b$, the $W$ boson can be produced in three helicity states: longitudinal ($\lambda=0$), left-handed ($\lambda=-1$), and right-handed ($\lambda=+1$). 
The corresponding helicity fractions are defined as
\begin{equation}
F_\lambda = \frac{\Gamma_\lambda}{\Gamma_{\text{tot}}}, \qquad \lambda = L, R, 0,
\end{equation}
where $\Gamma_\lambda$ denotes the partial decay width for a given $W$-boson helicity, and the total width is given by $\Gamma_{\text{tot}} = \Gamma_L + \Gamma_R + \Gamma_0$. 
By construction, the helicity fractions satisfy the normalization condition
$F_0 + F_L + F_R = 1$. At tree level in the SM, and neglecting the $b$-quark mass, these fractions are given by \cite{beneke, Bernreuther:2008ju ,Czarnecki:2010gb}
\begin{equation}
F_L = \frac{2 m_W^2}{m_t^2 + 2 m_W^2}, \qquad
F_R = 0, \qquad
F_0 = \frac{m_t^2}{m_t^2 + 2 m_W^2},
\end{equation}
where $m_t$ and $m_W$ denote the top-quark and $W$-boson masses, respectively. For $W^{-}$ bosons, 
the left- and right-handed helicity components are interchanged.
The vanishing of $F_R$ in the limit $m_b \to 0$ is a direct consequence of the purely left-handed structure of the charged 
weak current in the SM, together with angular momentum conservation. Finite $b$-quark mass effects and higher-order 
corrections generate a small but non-zero value of $F_R$, which remains strongly suppressed.
The relatively large value of the longitudinal fraction $F_0$ reflects the fact that the top quark is significantly 
heavier than the electroweak scale, and is closely related to the Goldstone boson equivalence theorem. 
As a result, top-quark decays provide a unique probe of the electroweak symmetry-breaking sector through the 
coupling of fermions to the longitudinal component of the $W$ boson.

The helicity fractions are modified in the presence of higher-dimensional operators in the SMEFT framework. 
At leading order in the EFT expansion, these effects arise from both dimension-6 and dimension-8 operators, 
which contribute through interference with the SM amplitude as well as through quadratic terms. 
Their impact on the helicity structure of the $Wtb$ vertex can be understood in terms of the induced Lorentz structures:
\begin{itemize}
\item \textbf{Vector operators} ($Q_{Hq}^{(3)}$, $Q_{Hud}^{(3)}$, $Q_{q^2H^4D}^{(2)}$, $Q_{udH^4D}$): 
These operators modify the vector structure of the $Wtb$ vertex. 
In particular, $Q_{Hq}^{(3)}$ and $Q_{q^2H^4D}^{(2)}$ correct the SM left-handed coupling, while 
$Q_{Hud}^{(3)}$ and $Q_{udH^4D}$ induce right-handed vector currents. 
The right-handed vector operators can lead to observable modifications of the helicity fractions, whereas the effects of 
$Q_{Hq}^{(3)}$ and $Q_{q^2H^4D}^{(2)}$ largely cancel in the ratios defining the helicity fractions, resulting in approximate flat directions in the fit.

\item \textbf{Dipole (tensor) operators} ($Q_{uW}$, $Q_{dW}$, $Q_{quWH^3}^{(1)}$, $Q_{qdWH^3}^{(1)}$): 
These operators introduce tensor structures in the $Wtb$ interaction, absent in the SM at tree level. 
They lift the helicity suppression of the right-handed $W$ configuration and can generate a non-zero $F_R$, while also modifying 
$F_L$ and $F_0$ through interference effects. The dimension-8 dipole operators contribute at $\mathcal{O}(\Lambda^{-4})$, 
the same order as the squared dimension-6 terms retained in our analysis, and can therefore affect the extraction of dimension-6 coefficients through additional correlations and degeneracies in the fit.
\end{itemize}
The complete analytical expressions for $\Gamma_{\text{tot}}$, $F_L$, $F_0$, and $F_R$, including a subset of 
dimension-6 and dimension-8 contributions, are provided in Appendix~\ref{app:decaywidths}.
Although we compute the total decay width $\Gamma_{\text{tot}}$ together with the helicity fractions for completeness and as part of the theoretical description of the decay, 
it is not included as an input observable in the statistical fit performed in this work. 
The fit is constructed only from the experimentally combined measurements of $F_L$ and $F_0$, for which a covariance matrix is available. 
The decay width is therefore used here to define the helicity fractions and to provide a consistent theoretical characterization of the $t \to Wb$ process, 
but not as an independent constraint in the extraction of Wilson coefficients.
Our numerical analysis uses the following input parameters taken from
the PDG~\cite{ParticleDataGroup:2024ppr}:
$m_W = 80.3692\,\text{GeV}$,
$m_b = 4.183\,\text{GeV}$,
$m_t = 172.57\,\text{GeV}$,
and $G_F = 1.166\times10^{-5}\,\text{GeV}^{-2}$.
Throughout this work we set $|V_{tb}| = 1$, neglecting effects from
other CKM elements. The EFT cutoff scale $\Lambda = 1$~TeV serves as
our reference value.

\subsection{Statistical Framework and Confidence Regions}

To derive constraints on the Wilson coefficients, we construct a $\chi^2$ function comparing theoretical predictions with experimental measurements:

\begin{equation}
\chi^2(\vec{C}) = \sum_{i,j=1}^2 [O_i^{\text{theo}}(\vec{C}) - O_i^{\text{exp}}] (\Sigma^{-1})_{ij} [O_j^{\text{theo}}(\vec{C}) - O_j^{\text{exp}}],
\end{equation}

where:
\begin{itemize}
\item $\vec{C} = \{C_1, C_2, \dots\}$ is the vector of Wilson coefficients
\item $O_i^{\text{theo}}(\vec{C})$ are theoretical predictions for observables $i = \{F_L, F_0\}$
\item $O_i^{\text{exp}}$ are experimental central values
\item $\Sigma$ is the covariance matrix
\end{itemize}

The covariance matrix is constructed from the combined ATLAS and CMS measurements of the W-boson helicity fractions~\cite{Aad:2020px}:
\begin{equation}
\Sigma = \begin{pmatrix}
\sigma_{F_0}^2 & \sigma_{F_0} \rho_{F_0 F_L} \sigma_{F_L}  \\
 \sigma_{F_L}\rho_{F_L F_0} \sigma_{F_0} & \sigma_{F_L}^2
\end{pmatrix}, 
\end{equation}
where the experimental uncertainties $\sigma_{F_0}$ and $\sigma_{F_L}$ are given in Table~\ref{tab:SMvsExp}, 
and the correlation coefficient $\rho_{F_0 F_L} = -0.85$ is taken from the same combined analysis~\cite{Aad:2020px}. 
This ensures statistical consistency between the central values and their correlations.

\noindent\textbf{(i) One-Parameter Fits:} For individual constraints, we perform one-parameter $\chi^2$ scans in which a single Wilson coefficient is varied while all other coefficients are fixed to zero:
\begin{equation}
\Delta\chi^2(C_k) = \chi^2(C_k;\, C_{j\neq k}=0) - \chi^2_{\min,k},
\end{equation}
where $\chi^2_{\min,k}$ is the minimum obtained in that one-parameter scan. The corresponding confidence intervals are determined using a $\chi^2$ distribution with one degree of freedom.

\noindent\textbf{(ii) Two-Parameter Fits:} To study correlations between operators, we perform selected two-parameter scans in which two Wilson coefficients are varied simultaneously while all remaining coefficients are fixed to zero:
\begin{equation}
\Delta\chi^2(C_k,C_l) = \chi^2(C_k,C_l;\, C_{m\neq k,l}=0) - \chi^2_{\min,kl},
\end{equation}
where $\chi^2_{\min,kl}$ is the minimum obtained in the corresponding two-parameter scan. This approach allows us to visualize degeneracies and correlations between different operator contributions, which is particularly important when dimension-6 and dimension-8 operators are considered simultaneously.


\section{Results}
\label{secres}

In this section, we present the numerical results of our analysis. 
We begin by establishing the SM baseline and comparing it with experimental measurements, 
and then derive constraints on the Wilson coefficients through one- and two-parameter $\chi^2$ fits.

\subsection{SM Baseline and Experimental Comparison}

As a foundation for our SMEFT analysis, we compute the SM predictions for the W-boson helicity fractions in the 
decay $t \to W b$, setting all higher-dimensional Wilson coefficients to zero, as presented in Appendix~\ref{app:decaywidths}. 
These theoretical predictions provide  the essential baseline against which we quantify SMEFT-induced deviations.

The experimental measurements, summarized in Table~\ref{tab:SMvsExp}, serve as the input for our statistical analysis. 
These results are taken from the combined ATLAS and CMS analysis~\cite{Aad:2020px}, 
which uses proton--proton collision data at $\sqrt{s}=8$~TeV with an integrated luminosity 
of about $20~\mathrm{fb}^{-1}$ per experiment. The measurements are based on 
top-quark pair events in the lepton+jets channel, where the $W$-boson helicity fractions 
are extracted from angular distributions of the decay products.

\begin{table}[H]
	\centering
	\renewcommand{\arraystretch}{1.5}
	\begin{tabular}{|c|c|}
		\hline
		\textbf{Observable} & \textbf{Experimental Value}  \\
		\hline
		$F_L$ & $0.315\pm 0.011$ \\
		$F_0$  & $0.693\pm 0.014$ \\
		\hline
	\end{tabular}
	\caption{Combined ATLAS and CMS measurements of the W-boson helicity fractions at $\sqrt{s}=8$~TeV~\cite{Aad:2020px}. 
	The same analysis provides the correlation $\rho_{F_0F_L} = -0.85$ used in the covariance matrix.}
	\label{tab:SMvsExp}
\end{table}

A combined likelihood fit is performed by the experimental collaborations, 
accounting for statistical and systematic uncertainties as well as their correlations. 
The fraction $F_R$ is not measured independently but obtained using the normalization 
constraint $F_0 + F_L + F_R = 1$.

For completeness, we note that more recent measurements of the $W$-boson helicity fractions have been reported by ATLAS using the full Run~2 dataset at $\sqrt{s}=13$~TeV with an integrated luminosity of $139~\mathrm{fb}^{-1}$~\cite{ATLAS:2022vfh}. The measured values are 
$F_0 = 0.684 \pm 0.005\,(\text{stat.}) \pm 0.014\,(\text{syst.})$, 
$F_L = 0.318 \pm 0.003\,(\text{stat.}) \pm 0.008\,(\text{syst.})$, 
and $F_R = -0.002 \pm 0.002\,(\text{stat.}) \pm 0.014\,(\text{syst.})$, 
in agreement with the SM prediction. 
The values used in our statistical analysis (Table~\ref{tab:SMvsExp}) correspond to the combined ATLAS and CMS result at $\sqrt{s}=8$~TeV, 
which provides a consistent set of central values, uncertainties, and correlations required for the construction of the covariance matrix.

This experimental input provides the basis for our $\chi^2$ analysis, allowing us to quantify deviations induced by dimension-6 and dimension-8 SMEFT operators.
Theoretical uncertainties, mainly associated with the omission of next-to-leading order QCD corrections, 
are expected to affect the predicted helicity fractions at the $1\text{--}2\%$ level~\cite{Zhang:2012cd, Czarnecki:2010gb, Fischer:2001gp}, 
and will be discussed in Sec.~\ref{secdis}.


\subsection{One-Parameter Constraints on Wilson Coefficients}

We derive constraints on individual Wilson coefficients through one-parameter $\chi^2$ scans, while setting all other coefficients to zero. 
Table~\ref{tab:WCinterval} reports the results only for those coefficients for which finite $68\%$ and $95\%$ confidence intervals are obtained at 
the benchmark scale $\Lambda = 1$ TeV. The remaining coefficients, namely $C_{Hq}^{(3)}$, $C_{q^2H^4D}^{(2)}$, and $C_{q^2H^4D}^{(3)}$, 
are not included in the table: $C_{q^2H^4D}^{(3)}$ does not contribute to the CP-even observables at $\mathcal{O}(\Lambda^{-4})$, 
while $C_{Hq}^{(3)}$ and $C_{q^2H^4D}^{(2)}$ correspond to flat directions in the one-parameter scans and therefore do not yield finite 
confidence intervals within the explored parameter range.

\begin{table}[H]
\centering
\renewcommand{\arraystretch}{1.5}
\begin{tabular}{|c|c|c|c|}
\hline
\textbf{Coefficient} & \textbf{Best Fit} & \textbf{68\% CL Range} & \textbf{95\% CL Range} \\
\hline
$C_{Hud}^{(3)}$ & $0.57$ & $[-1.84,\,2.99]$ & $[-3.77,\,4.96]$ \\
$C_{uW}$ & $0.23$ & $[0.05,\,0.42]$ & $[-0.13,\,0.60]$ \\
$C_{dW}$ & $0.17$ & $[-0.30,\,0.63]$ & $[-0.66,\,1.00]$ \\
\hline
$C_{quWH^3}^{(1)}$ & $7.47$ & $[1.53,\,13.41]$ & $[-4.17,\,19.11]$ \\
$C_{qdWH^3}^{(1)}$ & $98.81$ & $[31.53,\,166.09]$ & $[-33.03,\,230.65]$ \\
$C_{udH^4D}$ & $257.02$ & $[40.78,\,473.26]$ & $[-166.72,\,680.76]$ \\
\hline
\end{tabular}
\caption{One-parameter constraints on the subset of dimension-6 and dimension-8 Wilson coefficients for which 
finite $68\%$ and $95\%$ C.L. intervals are obtained at $\Lambda = 1$ TeV. Best-fit values are obtained from one-parameter 
$\chi^2$ scans with all other coefficients fixed to zero.}
\label{tab:WCinterval}
\end{table}

\begin{table}[H]
\centering
\renewcommand{\arraystretch}{1.5}
\begin{tabular}{|c|c|c|c|c|c|}
\hline
\makecell{Dim-6 Coeff.} &
\makecell{Dim-6 only (1D , 95\% CL)} &
$C_{quWH^3}^{(1)}$ &
$C_{qdWH^3}^{(1)}$ &
$C_{q^2 H^4 D}^{(2)}$ &
$C_{ud H^4 D}$ \\
\hline
$C_{uW}$
& [-0.13,\,0.60]
& [-0.50,\,1.03]
& [-0.29,\,0.64]
& [-0.21,\,0.69]
& [-0.22,\,0.68] \\
$C_{dW}$
& [-0.66,\,1.00]
& [-0.91,\,1.25]
& [-1.22,\,1.56]
& [-0.81,\,1.15]
& [-1.10,\,1.44] \\
$C_{Hud}^{(3)}$
& [-3.77,\,4.96]
& [-5.53,\,7.59]
& [-6.85,\,7.79]
& [-4.61,\,5.82]
& [-5.96,\,6.85] \\
\hline
\end{tabular}
\caption{Comparison of 95\% confidence intervals for selected dimension-6 Wilson coefficients. The second column shows the one-parameter dimension-6-only result with all dimension-8 coefficients fixed to zero. The remaining columns show the projected 95\% intervals obtained from two-parameter fits in which one dimension-6 coefficient and one dimension-8 coefficient are varied simultaneously, with all other coefficients set to zero.}
\label{tab:comparison_dim6_dim8}
\end{table}

To explicitly assess the impact of dimension-8 contributions on the extraction of dimension-6 coefficients, we compare the constraints 
obtained in a dimension-6-only fit (with all dimension-8 coefficients set to zero) to those derived from two-parameter fits in which one 
dimension-6 and one dimension-8 coefficient are varied simultaneously. For each case, the resulting $95\%$ confidence region is projected 
onto the dimension-6 axis. The results are presented in Table~\ref{tab:comparison_dim6_dim8}.
We observe that the inclusion of dimension-8 operators modifies the allowed ranges of the dimension-6 coefficients. 
In general, the projected constraints become weaker and, in some cases, more asymmetric when dimension-8 effects are included. 
This behavior reflects the presence of additional correlations and approximate degeneracies in the extended parameter space. 
These results demonstrate explicitly that neglecting dimension-8 contributions can lead to an incomplete or biased inference of dimension-6 effects.

The extracted confidence intervals are not necessarily symmetric around the best-fit values due to the non-linear dependence of 
the observables on the Wilson coefficients. In particular, the inclusion of $\mathcal{O}(\Lambda^{-4})$ contributions, 
arising from squared dimension-6 terms and SM--dimension-8 interference, leads to non-parabolic $\chi^2$ profiles. 
As a result, the likelihood can exhibit asymmetries, especially in directions with limited sensitivity, which in turn produces asymmetric confidence intervals.

As a validation of our analysis framework, we compare our dimension-6
constraints with those obtained in the combined ATLAS and CMS analysis
of $W$-boson polarization measurements in top-quark decays
\cite{Aad:2020px}.
For the dipole operator $C_{uW}$ (denoted $C_{tW}$
in Ref.~\cite{Aad:2020px}), our $95\%$ CL interval $[-0.125,\,0.598]$
is broadly compatible with the combined result $[-0.48,\,0.29]$.
Similarly, our limits on $C_{dW}$ ($[-0.659,\,0.996]$) and
$C_{Hud}^{(3)}$ ($[-3.769,\,4.963]$) are comparable to the corresponding
combined bounds $[-0.96,\,0.67]$ and $[-3.48,\,5.16]$, respectively.
The remaining differences arise from the different analysis strategies:
the ATLAS+CMS combination performs a bin-by-bin likelihood fit to the
differential decay distributions (notably the $\cos\theta^\ast$
distribution), whereas our analysis uses the experimentally extracted
helicity fractions $F_i$ as input observables. The inclusion of full
shape information in the experimental combination provides additional
sensitivity and therefore leads to tighter constraints. Overall, the
agreement in the allowed parameter ranges provides a useful validation
of our implementation.

Several important observations emerge from these results. The dipole operator $C_{uW}$ exhibits particularly narrow confidence intervals, 
indicating strong sensitivity driven by its direct modification of the tensor structure of the $Wtb$ vertex and its 
pronounced impact on the right-handed helicity fraction $F_R$. Most dimension-6 operators display intermediate sensitivity, 
with 95\% C.L. bounds typically of $\mathcal{O}(1)$ for $\Lambda = 1$ TeV. Coefficients such as $C_{Hq}^{(3)}$ 
show no real solution for the confidence intervals within the explored parameter range, indicating the presence of flat directions in the one-parameter fit.
More precisely, in these cases the $\chi^2$ function does not reach a minimum for the corresponding confidence level 
within the scanned parameter domain, preventing the extraction of statistically meaningful intervals. 
In contrast, dimension-8 operators are generally weakly constrained, with allowed ranges extending up to $\mathcal{O}(10^3)$, depending on the operator.
For some coefficients, in particular $C_{q^2H^4D}^{(2)}$, no finite confidence intervals can be extracted due to flat directions in the $\chi^2$ landscape. 

The relatively large best-fit values observed for some dimension-8 operators do not reflect a physical hierarchy in the EFT expansion, 
but rather the limited sensitivity of the helicity observables to these coefficients. In particular, several dimension-8 directions are only 
weakly constrained or exhibit approximate flat directions, allowing for large variations in the corresponding Wilson coefficients without significantly affecting the observables.
Overall, these findings highlight both the complementarity 
and the limitations of single-coefficient fits, suggesting that correlated multi-parameter analyses and improved experimental precision will be necessary to lift the remaining flat directions.
This behavior indicates that the current dataset does not provide strong constraints on these operators, and that additional observables will be required to probe the corresponding directions in parameter space more effectively.

These results underscore the complementarity between dimension-6 and dimension-8 operators and emphasize the need for future precision data, 
particularly from HL-LHC, to improve constraints on the poorly bounded coefficients.

\subsection{Correlated Constraints from Two-Parameter Fits}

\begin{figure}[ht!]
  \centering
  \begin{subfigure}{0.32\textwidth}
    \includegraphics[width=\linewidth]{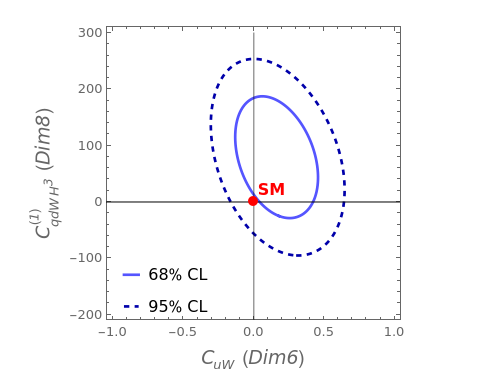}
  \end{subfigure}
  \begin{subfigure}{0.32\textwidth}
    \includegraphics[width=\linewidth]{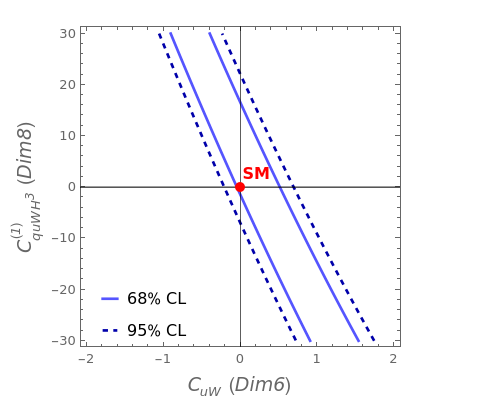}
  \end{subfigure}
    \\
  \begin{subfigure}{0.32\textwidth}
    \includegraphics[width=\linewidth]{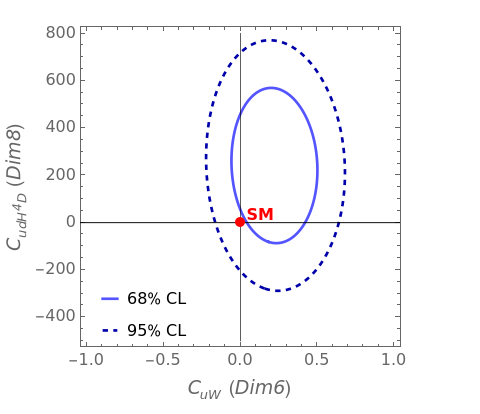}
  \end{subfigure}
  \caption{Confidence contours for $C_{uW}$ paired with dimension-8 operators.
   dashed lighter blue : 68\% CL; Solid Darker blue: 95\% CL.}
  \label{fig:contours_uW0}
\end{figure}

\begin{figure}[ht!]
  \centering
  \begin{subfigure}{0.32\textwidth}
    \includegraphics[width=\linewidth]{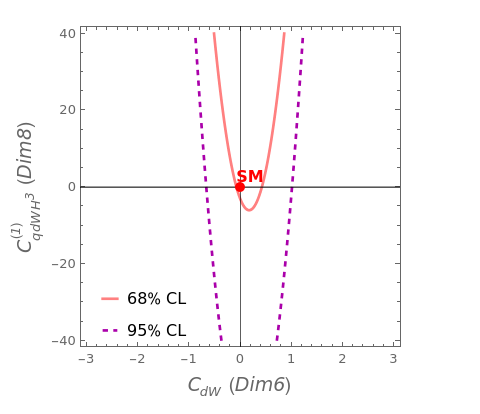}
  \end{subfigure}
  \begin{subfigure}{0.32\textwidth}
    \includegraphics[width=\linewidth]{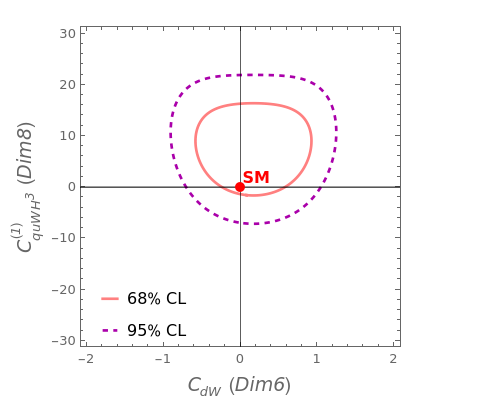}
  \end{subfigure}
  \\
  \begin{subfigure}{0.32\textwidth}
    \includegraphics[width=\linewidth]{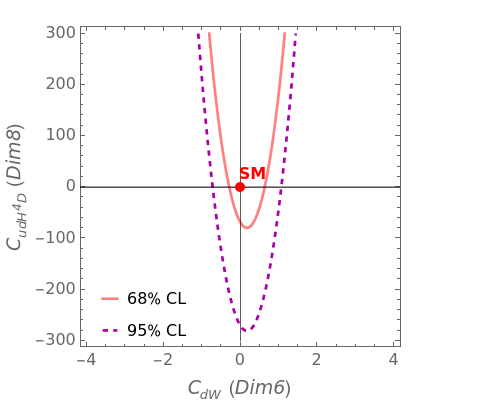}
  \end{subfigure}
  \caption{Confidence contours for $C_{dW}$ paired with dimension-8 operators.
   dashed pink : 68\% CL; Solid magenta: 95\% CL}
  \label{fig:contours_dW0}
\end{figure}

\begin{figure}[ht!]
  \centering
  \begin{subfigure}{0.344\textwidth}
    \includegraphics[width=\linewidth]{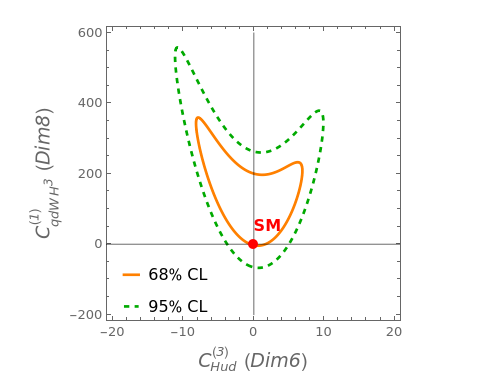}
  \end{subfigure}
  \begin{subfigure}{0.32\textwidth}
    \includegraphics[width=\linewidth]{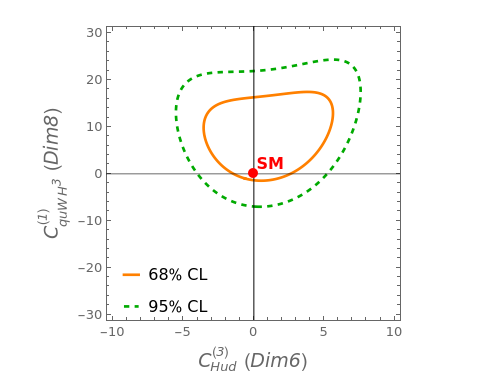}
  \end{subfigure}
    \\
  \begin{subfigure}{0.32\textwidth}
    \includegraphics[width=\linewidth]{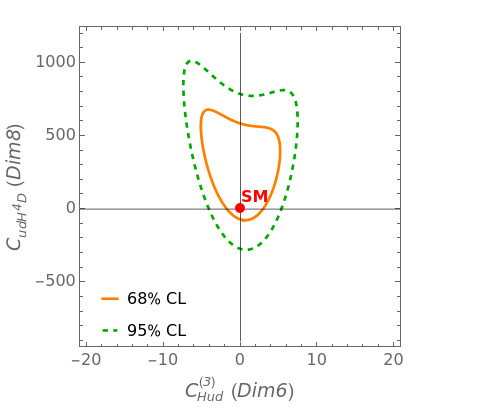}
  \end{subfigure}
  \caption{Confidence contours for $C_{Hud}^{(3)}$ paired with dimension-8 operators.
  dashed orange : 68\% CL; Solid green : 95\% CL.}
  \label{fig:contours_Hud0}
\end{figure}

To visualize correlations and degeneracies between operators of different mass dimensions, we perform 
selected two-parameter $\chi^2$ scans and present the resulting confidence regions in Figs.~\ref{fig:contours_uW0}--\ref{fig:contours_Hud0}. 
These planes involve representative combinations of dimension-6 and dimension-8 coefficients and illustrate how the inclusion of $\mathcal{O}(\Lambda^{-4})$ terms modifies the structure of the allowed parameter space.
The resulting confidence regions display a variety of characteristic shapes, reflecting the interplay between different operator structures in the effective $Wtb$ vertex. Compact closed contours correspond to genuine simultaneous constraints on both coefficients with moderate correlations, whereas elongated or open regions indicate approximate degeneracy directions, where only particular linear combinations of Wilson coefficients are constrained by the helicity observables.

In Fig.~\ref{fig:contours_uW0}, which shows planes involving the dipole coefficient $C_{uW}$, both types of behavior are observed. The $(C_{uW},\,C_{qdWH^3}^{(1)})$ and $(C_{uW},\,C_{udH^4D})$ planes exhibit closed confidence regions, indicating nontrivial correlations but still meaningful sensitivity to both coefficients. By contrast, the $(C_{uW},\,C_{quWH^3}^{(1)})$ plane is represented by a narrow open band rather than a closed contour. This signals a strong anti-correlation and an approximate degeneracy between the two coefficients. Such behavior is expected because both operators contribute to the same tensor structure of the $Wtb$ vertex, so their effects on the helicity fractions can compensate each other over an extended region of parameter space.

A different pattern is observed in Fig.~\ref{fig:contours_dW0}. The $(C_{dW},\,C_{qdWH^3}^{(1)})$ and $(C_{dW},\,C_{udH^4D})$ planes show open, 
parabolic-shaped confidence regions, indicating non-linear degeneracies between dimension-6 and dimension-8 contributions. 
These features arise from the interplay between squared dimension-6 terms and SM--dimension-8 interference at $\mathcal{O}(\Lambda^{-4})$, 
which allows the dimension-8 contributions to compensate the quadratic dependence on the dimension-6 coefficient. 
In contrast, the $(C_{dW},\,C_{quWH^3}^{(1)})$ plane displays a closed and approximately elliptical contour, 
indicating genuine simultaneous sensitivity to both coefficients with only moderate correlations. 
This demonstrates that the structure of the constraints depends strongly on the Lorentz nature of the operators involved and that not all dimension-6/dimension-8 combinations generate comparable degeneracies.

The most pronounced non-Gaussian features appear in Fig.~\ref{fig:contours_Hud0}, involving the right-handed vector coefficient $C_{Hud}^{(3)}$. In particular, the $(C_{Hud}^{(3)},\,C_{qdWH^3}^{(1)})$ plane shows a strongly distorted and cusp-like allowed region, indicating significant non-linear correlations in the $\chi^2$ function. This behavior results from the interplay between the right-handed vector contribution at dimension~6 and the tensor structure induced by the dimension-8 operator, leading to a clearly non-quadratic dependence of the observables on the Wilson coefficients. The $(C_{Hud}^{(3)},\,C_{quWH^3}^{(1)})$ plane exhibits a smoother but still asymmetric closed contour, corresponding to a moderate correlation with residual non-linear effects. 
Finally, the $(C_{Hud}^{(3)},\,C_{udH^4D})$ plane is strongly elongated and asymmetric, indicating unequal sensitivity to the two coefficients, with the dimension-8 coefficient remaining only weakly constrained.

Across all planes, the SM point $(C_i=0)$ lies within the 95\% confidence regions, indicating overall consistency with current experimental data. At the same time, in several planes the preferred regions are displaced from the origin, in qualitative agreement with the nonzero best-fit values obtained in the one-parameter scans. This shows that the inclusion of dimension-8 contributions modifies both the location and the geometry of the allowed parameter space, even in the absence of a statistically significant deviation from the SM.

Taken together, these results show that the impact of dimension-8 operators is non-uniform. In some directions they induce only mild shifts in the allowed regions, while in others they generate strong correlations, 
extended degeneracies, or manifestly non-elliptic confidence regions. The contour plots therefore provide important information beyond that contained in one-parameter limits alone.
 A more complete SMEFT determination would require the inclusion of additional observables sensitive to the same interaction, such as single-top production, top-pair production and decay distributions, 
 and electroweak precision observables. Such a combined analysis would help lift degeneracies and provide a more global determination of the relevant SMEFT parameter space.

All results presented in this work are obtained within a consistent EFT expansion up to $\mathcal{O}(\Lambda^{-4})$, 
including SM--dimension-6 interference, squared dimension-6 terms, and SM--dimension-8 interference. 
We have verified that the inclusion of these $\mathcal{O}(\Lambda^{-4})$ contributions modifies the 
inferred constraints in a controlled and physically meaningful way, confirming the consistency of the adopted EFT expansion scheme in the present analysis.

\section{Discussion}
\label{secdis}

The results presented in this work highlight the importance of consistently including dimension-8 operators in SMEFT analyses of top-quark decays. 
At $\mathcal{O}(\Lambda^{-4})$, squared dimension-6 contributions and SM--dimension-8 interference enter at the same order, 
and therefore must be treated on equal footing. Neglecting dimension-8 effects while retaining quadratic dimension-6 terms 
leads to an incomplete description of the parameter space and can bias the interpretation of the extracted constraints.

The variation in sensitivity among the different operators reflects their underlying Lorentz structures.
In particular, the strong constraint on $C_{uW}$ arises from its ability to induce a right-handed $W$ helicity component, 
which is strongly suppressed in the SM in the $m_b \to 0$ limit. 
Although $F_R$ is not included as an independent observable in the fit, its effects are indirectly probed through the normalization condition 
$F_0 + F_L + F_R = 1$, such that any enhancement of the right-handed contribution necessarily modifies the measured fractions $F_0$ and $F_L$. 
In contrast, operators modifying the dominant left-handed structure exhibit weaker bounds and, in some cases, approximate flat directions, 
reflecting the limited resolving power of helicity observables in the presence of multiple operator contributions.

The two-parameter fits further demonstrate that correlations and degeneracies between operators are a generic feature of the EFT description. 
Different operator combinations can lead to similar predictions for the helicity fractions, resulting in elongated or non-elliptic confidence regions. 
This behavior reflects the intrinsically non-Gaussian structure of the likelihood once $\mathcal{O}(\Lambda^{-4})$ contributions are included, 
and illustrates the limitations of approximating the parameter space with a simple quadratic expansion.

Theoretical uncertainties associated with missing higher-order QCD corrections are expected to affect the helicity fractions at the percent level \cite{Czarnecki:2010gb}. 
While these effects ($1-2\%$) are smaller than current experimental uncertainties, they will become  relevant for future precision studies. 
NLO SMEFT corrections may also modify the quantitative interpretation of the constraints, although they are not expected to alter 
the qualitative features observed in this work.

Overall, the results indicate that helicity fractions alone are not sufficient to fully constrain the operator space. 
A comprehensive SMEFT analysis will require the inclusion of complementary observables, such as single-top production, 
top-pair processes, and differential decay distributions, which probe different combinations of operator structures. 
Future measurements at the HL-LHC will play a crucial role in lifting degeneracies and improving sensitivity to higher-dimensional interactions.


\section{Conclusion}
\label{secsummary}

We have performed a combined SMEFT analysis of dimension-6 operators together with a representative subset of dimension-8 contributions in top-quark decays, using the $W$-boson helicity fractions as precision observables. 
Working consistently at $\mathcal{O}(\Lambda^{-4})$, we included both squared dimension-6 terms and SM--dimension-8 interference.

The resulting constraints exhibit a clear difference in sensitivity among the various operators.
The dipole operator $C_{uW}$ is the most tightly constrained, with a $95\%$ CL interval of $[-0.125,\,0.598]$, 
reflecting the strong sensitivity of the suppressed right-handed helicity fraction $F_R$ to tensor interactions. 
In contrast, operators affecting the dominant left-handed $Wtb$ structure are more weakly constrained, 
with $C_{dW}$ and $C_{Hud}^{(3)}$ bounded at the $\mathcal{O}(1)$ level, while 
dimension-8 coefficients are generally weakly constrained, with allowed ranges extending up 
to $\mathcal{O}(10^3)$ or larger, depending on the operator and the presence of degeneracies in the parameter space.

The two-parameter fits reveal non-trivial correlations and degeneracies between dimension-6 and dimension-8 operators, 
leading to elongated or non-elliptic confidence regions. 
These features demonstrate that higher-dimensional contributions modify not only the numerical bounds but also the geometry of the allowed parameter space. 
Since squared dimension-6 terms and SM--dimension-8 interference enter at the same order, 
neglecting dimension-8 effects while retaining quadratic dimension-6 contributions leads to an incomplete EFT interpretation.

Finally, the observed degeneracies indicate that helicity fractions alone are insufficient to fully constrain the operator space. 
Future improvements, particularly at the HL-LHC, together with complementary observables such as single-top and top-pair processes, 
will be essential to lift these degeneracies and achieve a more complete determination of the $Wtb$ interaction.

\section*{Acknowledgement}
We would like to especially thank Dr.~Seddigheh Tizchang for valuable discussions.


\appendix
\section{Total Decay Width}
\label{app:decaywidths}

In this appendix, we provide the detailed derivation of the top-quark decay width 
and the helicity fractions of the $W$ boson corresponding to the longitudinal, 
left-handed, and right-handed polarization states. 
We present analytic expressions for the contributions arising from dimension-6 
and dimension-8 SMEFT operators, together with the relevant interference terms 
that enter the $\chi^2$ analysis performed in the main text.
The full squared amplitudes presented below include all operator structures for completeness. 
However, in the numerical analysis performed in the main text, terms involving CP-odd operators and contributions 
beyond $\mathcal{O}(\Lambda^{-4})$ are consistently neglected.

The total decay width of the top quark in the SMEFT framework can be expressed as
an expansion in inverse powers of the new-physics scale $\Lambda$:
\begin{equation}
	\Gamma(t \to W b) \;=\; \Gamma_{\rm SM} 
	+ \frac{1}{\Lambda^2} \, \Gamma_{\rm D6}^{\rm int} 
	+ \frac{1}{\Lambda^4} \, \Gamma^{2}_{\rm D6} 
	+ \frac{1}{\Lambda^4} \, \Gamma_{\rm D8}^{\rm int} 
	+ \dots \, ,
	\label{eq:Gamma_total}
\end{equation}
where:
\begin{itemize}
	\item $\Gamma_{\rm SM}$ denotes the Standard Model contribution.
	\item $\Gamma_{\rm D6}^{\rm int}$ represents the interference between the SM amplitude and amplitudes induced by dimension-6 operators.
	\item $\Gamma^{2}_{\rm D6}$ corresponds to the squared contribution of dimension-6 operators.
	\item $\Gamma_{\rm D8}^{\rm int}$ denotes the interference between the SM and dimension-8 operators.
\end{itemize}
The ellipsis indicates higher-order terms in the SMEFT expansion, which are neglected 
throughout this work.

All expressions (including the complete squared amplitudes,
intermediate steps, and Mathematica notebooks used to generate the results)
are now provided as supplementary material and are publicly available at:

\begin{center}
\url{https://github.com/afsanehkianfar/Paper-SMEFT-twb}
\end{center}

These files contain the full unshortened expressions and allow complete reproducibility of all numerical results presented in this work.

\section{W-boson Helicity Fractions}
\label{app:WHelicity}

The SM  predictions for the W boson helicity fractions are defined as:
\begin{align}
	F_L^{\text{SM}} &= \frac{m_W^2 \left( \sqrt{-(m_b - m_t - m_W)(m_b + m_t - m_W)} 
		\sqrt{m_t^2 - (m_b + m_W)^2} + m_b^2 + m_t^2 - m_W^2 \right)}
	{m_W^2(m_b^2 + m_t^2) + (m_b^2 - m_t^2)^2 - 2m_W^4}, \nonumber \\[6pt]
	F_R^{\text{SM}} &= \frac{m_W^2 \left( -\sqrt{-(m_b - m_t - m_W)(m_b + m_t - m_W)} 
	\sqrt{m_t^2 - (m_b + m_W)^2} + m_b^2 + m_t^2 - m_W^2 \right)}
	{m_W^2(m_b^2 + m_t^2) + (m_b^2 - m_t^2)^2 - 2m_W^4}, \nonumber \\[6pt]
		F_0^{\text{SM}} &= \frac{(m_b^2 - m_t^2)^2 - m_W^2(m_b^2 + m_t^2)}
	{m_W^2(m_b^2 + m_t^2) + (m_b^2 - m_t^2)^2 - 2m_W^4}.
\end{align}

We present here the analytic results for the total decay width $\Gamma_{\textrm{tot}}$ of $t \rightarrow W b$ at leading order (LO) in SMEFT. 
The intermediate expressions shown below correspond to the full squared amplitude,
which formally contains terms up to $\mathcal{O}(\Lambda^{-8})$.
For the phenomenological analysis presented in the main text,
the EFT expansion is consistently truncated at $\mathcal{O}(\Lambda^{-4})$,
and higher-order terms are discarded.
\allowdisplaybreaks

\begin{align}
	\Gamma_{\text{tot}} &= 
	\frac{1}{4096 \, \Lambda^8 \, m_t^7 \, m_W^2 \, \pi}\,
	\sqrt{(m_b + m_t - m_W)(-m_b + m_t + m_W)}\,
	\sqrt{m_t^2 - (m_b + m_W)^2}\, N_c
	\nonumber 
	\\[6pt]&\qquad
	\times \Bigg(16 g^2 \Lambda^4 m_t^4 \Big[(m_b^2 - m_t^2)^2 + (m_b^2+m_t^2)m_W^2 -2m_W^4\Big] v^4 
	\,[C_{Hud}^{(3)*}]^2
	\nonumber 
	\\[6pt]&\qquad
	+ 16 m_t^4 v^2 \Big(
	-8 m_W^2 \big[-2(m_b^2-m_t^2)^2+(m_b^2+m_t^2)m_W^2+m_W^4\big] v^4 \, [C_{qdWH^3}^{(1)*}]^2
	\nonumber 
	\\[6pt]&\qquad
	-32 \Lambda^4 m_W^2 \big[-2(m_b^2-m_t^2)^2+(m_b^2+m_t^2)m_W^2+m_W^4\big][C^*_{dW}]^2
	\nonumber 
	\\[6pt]&\qquad
	-24\sqrt{2}\, g \Lambda^2 m_t m_W^2 (m_b^2-m_t^2+m_W^2)v^3
	\,[C^*_{dW}]\,[C^*_{udH^4D}]
	\nonumber 
	\\[6pt]&\qquad
	+ g^2 \big[(m_b^2-m_t^2)^2+(m_b^2+m_t^2)m_W^2-2m_W^4\big] v^6 [C^*_{udH^4D}]^2
	\nonumber 
	\\[6pt]&\qquad
	-4 m_W^2 v^2 [C_{qdWH^3}^{(1)*}]
	\Big( 8\Lambda^2 \big[-2(m_b^2-m_t^2)^2+(m_b^2+m_t^2)m_W^2+m_W^4\big][C^*_{dW}]
	\nonumber 
	\\[6pt]&\qquad
	+3\sqrt{2} g m_t (m_b^2-m_t^2+m_W^2) v^3 [C^*_{udH^4D}]\Big)\Big)
	-32 g \Lambda^2 m_t^4 v^2 [C_{Hud}^{(3)*}]
	\nonumber 
	\\[6pt]&\qquad
	\times \Big(6\sqrt{2} m_t m_W^2 (m_b^2-m_t^2+m_W^2)v^3 [C_{qdWH^3}^{(1)*}]
	+12\sqrt{2} \Lambda^2 m_t m_W^2 (m_b^2-m_t^2+m_W^2) v
	\nonumber 
	\\[6pt]&\qquad
	 \times [C^*_{dW}]
	- g\big[(m_b^2-m_t^2)^2+(m_b^2+m_t^2)m_W^2-2m_W^4\big]v^4 [C^*_{udH^4D}]
	\nonumber 
	\\[6pt]&\qquad
	+6 m_b m_W^2 [V^*_{tb}]
	\Big(-\sqrt{2}(m_b^2-m_t^2-m_W^2)v\,(2C_{uW}\Lambda^2+C_{quWH^3}^{(1)} v^2)
	\nonumber 
	\\[6pt]&\qquad
	+ g m_t \big(2\Lambda^4+(C_{q^2H^4D}^{(2)}+C_{q^2H^4D}^{(3)})v^4\big)
	+ g m_t v^2 \big(2\Lambda^2 [C_{Hq}^{(3)*}]
	\nonumber 
	\\[6pt]&\qquad
	+ v^2 ([C_{q^2H^4D}^{(2)*}] - \Re[C_{q^2H^4D}^{(3)}])\big)\Big)\Big)
	\nonumber \\[6pt]&\qquad
	+ 48 m_b m_t \Big(m_b^2-m_t^2-m_W^2 
	-\sqrt{(m_b+m_t-m_W)(-m_b+m_t+m_W)}\sqrt{m_t^2-(m_b+m_W)^2}\Big)
	\nonumber \\[6pt]&\qquad
	\times \Big(m_b^2-m_t^2-m_W^2 
	+\sqrt{(m_b+m_t-m_W)(-m_b+m_t+m_W)}\sqrt{m_t^2-(m_b+m_W)^2}\Big)
	\nonumber \\[6pt]&\qquad
	\times v \, [V^*_{tb}]
	\Big(g m_t v^3 [C^*_{udH^4D}]
	\big(\sqrt{2}(m_b^2-m_t^2-m_W^2) v(2C_{uW}\Lambda^2+C_{quWH^3}^{(1)}v^2)
	\nonumber \\[6pt]&\qquad
	-2g \Lambda^2 m_t v^2 [C_{Hq}^{(3)*}]-g m_t \Big(2\Lambda^4 + C_{q^2H^4D}^{(3)}v^4 
	\nonumber \\[6pt]&\qquad
    + v^4(C_{q^2H^4D}^{(2)}+[C_{q^2H^4D}^{(2)*}]-\Re[C_{q^2H^4D}^{(3)}])\Big)
	\nonumber \\[6pt]&\qquad
	- m_t (v^2 [C_{qdWH^3}^{(1)*}] + 2\Lambda^2 [C^*_{dW}])
	\big( 8 m_t m_W^2 v(2C_{uW}\Lambda^2+C_{quWH^3}^{(1)} v^2)
	\nonumber \\[6pt]&\qquad 
    -\sqrt{2} g (m_b^2-m_t^2-m_W^2)\big(2\Lambda^4+C_{q^2H^4D}^{(3)}v^4 +2\Lambda^2 v^2  [C_{Hq}^{(3)}]
	\nonumber \\[6pt]&\qquad
	+ v^4(C_{q^2H^4D}^{(2)}+[C_{q^2H^4D}^{(2)*}]-\Re[C_{q^2H^4D}^{(3)}])\big)\big)\Big)
	\nonumber \\[6pt]&\qquad
	+ [V^*_{tb}]^2 \Big(
	4 m_t^4 (m_b^2-m_t^2+m_W^2)
	\Big(-64 m_W^2 (-m_b^2+m_t^2+m_W^2) v^2 
	\nonumber \\[6pt]&\qquad
	\times (2C_{uW}\Lambda^2+C_{quWH^3}^{(1)}v^2)^2
	-48\sqrt{2} g m_t m_W^2 v (2C_{uW}\Lambda^2+C_{quWH^3}^{(1)}v^2)
	(2\Lambda^4+C_{q^2H^4D}^{(3)}v^4
	\nonumber \\[6pt]&\qquad
	+2\Lambda^2 v^2 [C_{Hq}^{(3)*}]
	+ v^4(C_{q^2H^4D}^{(2)}+[C_{q^2H^4D}^{(2)*}]-\Re[C_{q^2H^4D}^{(3)}]))
	\nonumber \\[6pt]&\qquad
	+ g^2 (m_b^2-m_t^2-m_W^2)\Big(
	16\Lambda^8+(4{C_{q^2H^4D}^{(2)}}^2+{C_{q^2H^4D}^{(3)}}^2)v^8
	+8C_{q^2H^4D}^{(2)}v^8 [C_{q^2H^4D}^{(2)*}]
	\nonumber \\[6pt]&\qquad
	+4v^8 [C_{q^2H^4D}^{(2)*}]^2
	+16\Lambda^4 v^4 [C_{Hq}^{(3)*}]^2
	\nonumber \\[6pt]&\qquad
	+ v^8(-2 C_{q^2H^4D}^{(3)}[C_{q^2H^4D}^{(3)*}]
	+[C_{q^2H^4D}^{(3)*}]^2
	+16 i \Im[C_{q^2H^4D}^{(3)}]\Re[C_{q^2H^4D}^{(2)}])
	\nonumber \\[6pt]&\qquad
	+8\Lambda^2 v^2 [C_{Hq}^{(3)*}]
	\big(4\Lambda^4+2v^4(C_{q^2H^4D}^{(2)}+C_{q^2H^4D}^{(3)}+[C_{q^2H^4D}^{(2)*}]-\Re[C_{q^2H^4D}^{(3)}])\big)
	\nonumber \\[6pt]&\qquad
	+16 \Lambda^4 v^4 (C_{q^2H^4D}^{(2)}+C_{q^2H^4D}^{(3)}+[C_{q^2H^4D}^{(2)*}]-\Re[C_{q^2H^4D}^{(3)}])\Big)\Big)
	\nonumber \\[6pt]&\qquad
	- m_t^2(m_b^2+m_t^2-m_W^2)
	\Big(m_b^2-m_t^2-m_W^2
	-\sqrt{(m_b+m_t-m_W)(-m_b+m_t+m_W)}\sqrt{m_t^2-(m_b+m_W)^2}\Big)
	\nonumber \\[6pt]&\qquad
	\times 
	\Big(m_b^2-m_t^2-m_W^2
	+\sqrt{(m_b+m_t-m_W)(-m_b+m_t+m_W)}\sqrt{m_t^2-(m_b+m_W)^2}\Big)
	\nonumber \\[6pt]&\qquad
	\times \Big(32 m_W^2 v^2 (2C_{uW}\Lambda^2+C_{quWH^3}^{(1)}v^2)^2
	- g^2 (16\Lambda^8+16(C_{q^2H^4D}^{(2)}+C_{q^2H^4D}^{(3)})\Lambda^4 v^4
	\nonumber \\[6pt]&\qquad
	+(4{C_{q^2H^4D}^{(2)}}^2+{C_{q^2H^4D}^{(3)}}^2)v^8)
	-g^2 v^2\big(
	8 v^2(2\Lambda^4+C_{q^2H^4D}^{(2)}v^4)[C_{q^2H^4D}^{(2)*}]
	\nonumber \\[6pt]&\qquad
	+4 v^6 [C_{q^2H^4D}^{(2)*}]^2
	+16\Lambda^4 v^2 [C_{Hq}^{(3)*}]^2
	-2 C_{q^2H^4D}^{(3)} v^6 [C_{q^2H^4D}^{(3)*}]
	\nonumber \\[6pt]&\qquad
	+ v^6 [C_{q^2H^4D}^{(3)*}]^2
	+16 i v^6 \Im[C_{q^2H^4D}^{(3)}] \Re[C_{q^2H^4D}^{(2)}]
	+8\Lambda^2 [C_{Hq}^{(3)*}]
	\nonumber \\[6pt]&\qquad
	\times (4\Lambda^4+2v^4(C_{q^2H^4D}^{(2)}+C_{q^2H^4D}^{(3)}+[C_{q^2H^4D}^{(2)*}]-\Re[C_{q^2H^4D}^{(3)}]))
	-16\Lambda^4 v^2 \Re[C_{q^2H^4D}^{(3)}]
	\big)\Big)\Big)\Bigg).
\end{align}

\end{document}